\documentclass[onecolumn]{aastex6}

\newcounter{magicrownumbers}

\slugcomment{Submitted to The Astrophysical Journal}
\shorttitle{
}
\shortauthors{Sion et al.}
\watermark{Accepted version}
\setwatermarkfontsize{1.1in}

\begin{document}

\title{{\bf 
{\it Fuse} Spectroscopic Analysis of the Slowest Symbiotic Nova AG Peg During Quiescence      
}}

\author{Edward M. Sion\altaffilmark{1} } 
\author{Patrick Godon\altaffilmark{1,2} }
\author{Joanna Mikolajewska\altaffilmark{3} }
\author{Marcus Katynski\altaffilmark{1} }


\email{edward.sion@villanova.edu} 
\email{patrick.godon@villanova.edu}
\email{mikolaj@camk.edu.pl} 
\email{mkatynsk@villanova.edu} 

\altaffiltext{1}{Department of Astrophysics \& Planetary Science, 
Villanova University, Villanova, PA 19085, USA}
\altaffiltext{2}{Henry A. Rowland Department of Physics \& Astronomy,
The Johns Hopkins University, Baltimore, MD 21218, USA}
\altaffiltext{3}{N. Copernicus Astronomical Center, Polish Academy of Sciences, Bartycka 18, 00-716, Warsaw, Poland}

\begin{abstract}

We present a far ultraviolet spectroscopic analysis of the slowest 
known symbiotic nova AG Peg (M3/4III giant + hot white
dwarf; $P_{\rm orb} = 818.4$ days) which underwent a nova explosion in 1850 
followed by a very slow decline that did not end
until $\sim$1996, marking the beginning of quiescence. 
The $\sim$19 years of 
quiescence ended in June 2015, when AG Peg exhibited a Z And-type outburst with an optical amplitude of $\sim$1.5 magnitudes. 

We have carried out accretion disk and WD photosphere synthetic spectral 
modeling of a {\it Far Ultraviolet Spectroscopic Explorer 
(FUSE)} spectrum obtained on June 5.618, 2003 during 
the quiescence interval $\sim$12 years before the 2015 outburst. 
The spectrum is heavily affected by ISM absorption as well 
as strong emission lines. 
We de-reddened the {\it FUSE} fluxes assuming $E(B-V) = 0.10$, which is the
maximum galactic reddening in the direction of AG Peg. We discuss 
our adoption of the pre-Gaia distance over the Gaia parallax. 
For a range of white dwarf surface gravities and surface temperatures 
we find that the best-fitting photosphere is a hot WD with 
a temperature $T_{\rm wd}= 150,000$~K, and a low gravity $log(g) \sim 6.0-6.5$.
For a distance of 800 ~pc, the  scaled WD radius is 
$R_{\rm wd} \sim 0.06 \times R_{\odot}$, giving $log(g) = 6.67$ for 
a $0.65 M_{\odot}$ WD mass. 
The Luminosity we obtain from this 
model is $L=6.64 \times 10^{36}{\rm erg/s} = 1729 L_{\odot}$.   
The hot photosphere models provide better fits than the accretion disk models  
which have FUV flux deficits toward the shorter wavelengths of {\it FUSE}, 
down to the Lyman Limit. Given the uncertainty of the nature of a true symbiotic 
accretion disk, and, while a very hot low gravity degenerate star dominates the FUV flux, 
the presence of a steady-state (standard) accretion disk cannot be summarily ruled out.

\end{abstract}


\section{{\bf Introduction}} \label{sec:intro}



AG Peg is the slowest known symbiotic nova (M3/4III giant + 
hot white dwarf; $P_{\rm orb} = 818.4$ days) which underwent a nova 
explosion in 1850 followed by $\sim$165 years of decline and quiescence. 
A number of studies \citep{ken01,yoo06,yoo08,kim08,san17} document 
the very slow decline, since the 1850 nova explosion, and the 
transition into quiescence (up to $\sim$1995). 
From roughly 1997 to June 2015, the brightness of AG Peg  
appears to be fairly constant:   
its  visual magnitude (e.g. AAVSO light curve - LC) shows variation between 
about $\sim$9.3 and $\sim$8.3. Its photometric U-LC  clearly exhibits a modulation  
with the orbital phase \citep[][; attributed to the nebular component]{sko17}.  
This led \citet[][and references therein]{sko17} 
to confirm that AG Peg had at last reached its quiescence.  
In 2015, AG Peg unexpectedly transitioned to a weaker, sharper outburst 
with an optical amplitude of $\sim$1.5 magnitudes 
and short duration characteristic of Z And-type symbiotic variables. 
This new type of outburst may be due to gravitational 
energy release driven by (disk?) accretion and unlikely to be of 
thermonuclear origin like the 1850 outburst. This kind of transition 
of a symbiotic nova to Z And-type outbursts is unprecedented.

In order to shed light on the nature of the hot component (the 
"central engine"), we present here an analysis of a {\it FUSE} spectrum 
of AG Peg obtained in quiescence.  
In this far ultraviolet wavelength range down to the Lyman Limit,
the contribution of the nebular 
continuum produced by the photoionization of the red giant wind 
(which flattens the slope) is totally negligible. In this region, 
we are probing the flux of the innermost disk and white dwarf photosphere.

The published orbital and physical parameters of AG Peg are listed in
Table 1 with their references. 
Its Gaia \citep[DR2,][]{pru16,bro18} parallax is 
$\varpi \sim 0.380298.. \pm 0.081982..$ mas, 
giving a distance is $d = 2629_{-462}^{+725}$pc.
However, the standard uncertainty 
is $\sigma_{\varpi} \sim 0.02$ for a 9 mag star \citep[e.g. ][]{sch18} with such a parallax 
($\varpi \sim 0.4$ mas), or  four times smaller. 
The uncertainty in distance $( d = 1/\varpi)$ 
will have a subtantially non-Gaussian
shape when $\sigma_{\varpi}$ grows to a substantial fraction
of the parallax $\varpi$. The problem becomes non-trivial for cases
where $\sigma_{\varpi}/\varpi \sim 0.2$ or larger \citep{bai15}.  
The parallax of AG Peg has a significance of 4.6 sigma,  
and from its quality flags (e.g. goodness of fit, number of visibility periods) 
it appears that the Gaia parallax for AG Peg is actually unreliable
\citep[see e.g.][; 
see also 
https://gea.esac.esa.int/archive/documentation/GDR2/Data\_processing/ 
]{eye18,lur18}. 
One of the explanations for the unreliability of AG Peg parallax
is the fact that a long-period binary orbit will cause the 
center-of-light to wobble with a shift comparable to the parallaxe. 
As the binary separation is about 2~AU with a period of almost 2~yr,
over a 6~month period, the components move a distance of the order 
$\sim 1$~AU (depending on the exact mass ratio and inclination). 
 
Therefore, we mainly present results based on the widely 
accepted pre-Gaia distance to AG Peg of $\sim$800 pc.

\begin{deluxetable*}{lcl} 
\tablewidth{0pt}
\tablecaption{Orbital and Physical Parameters of AG Peg}
\tablehead{ 
Parameter        & Value        & References        
}
\startdata
Period                      &  818.4 d                          &    \citet{fek00}  \\ 
Inclination                 &  $50^{\circ}$                     &    \citet{ken93}  \\ 
Distance                    &  800 pc                           &    \citet{ken93}  \\
Separation                  &  $2.5 \pm 0.1$ AU                 &    \citet{ken93}  \\ 
$E(B-V)$                    &  0.10$\pm$0.05                    &    \citet{ken93}  \\ 
RG Spectral Type            &  M3/4 III                         &    \citet{ken93}  \\ 
WD Mass                     & $0.65  \pm 0.1 M_{\odot}$         &    \citet{ken93}  \\ 
WD Radius (quiescence)      & $\sim 0.05-0.06 R_{\odot}$        &    \citet{sko17}  \\ 
WD Temperature (quiescence) & 95k~K, $\sim$160k~K               &    \citet{mur91,sko17}  \\ 
RG Mass                     & $ 2.6 \pm 0.4 M_{\odot}$          &    \citet{ken93}  \\ 
RG Temperature              &  3,500~K                          &    \citet{ric99,vb99}   \\ 
\enddata
\end{deluxetable*}

In section 2, we provide the details of the 
{\it FUSE} observation including the complexities encountered and 
how we corrected for them, particularly the molecular hydrogen absorption 
that is pervasive in the {\it FUSE} wavelength range
along with other ISM absorption features (see subsection 2.1). 
In subsection 2.2, we describe our suite of modeling codes for both 
NLTE high gravity model atmospheres and model accretion disks:  
the \citet{wad98} grid of standard, steady state disk models and  
accretion disk models we have generated from scratch. 
Our grid has non-standard accretion disk models in the form of 
(inner and outer) disk truncation,
and covers parameter space outside of the Wade and Hubeny grid. 
In section 3, we present the results of our synthetic spectral analysis 
of the {\it FUSE} data using accretion disk models and white dwarf models. 
Finally, in section 4 we summarize our conclusions.

\section{Observations and Method of Analysis} 

\subsection{The {\it FUSE} Spectrum.}

AG Peg was observed with {\it FUSE} on June 5.618, 2003 
(exposure start time MJD 52795.61791667) and 
data from the AAVSO indicate it had a magnitude $m_v=8.5$.
The orbital phase at the time of the {\it FUSE} observation 
was $\phi = 0.31$, which we derived using the orbital ephemeris of 
\citet{fek00} 2 447 165.3($\pm$48)+ 818.2($\pm$1.6) x E  
(itself based upon the inferior conjunction of the M3/4 giant). 
We retrieved the {\it FUSE} data (ID Q1110103) from the MAST archive. 
The instrument was set in time-tag and the data were obtained  
through the MDRS aperture with an exposure time of 2116 seconds 
(one single {\it FUSE} orbit). 
The data were calibrated through the pipeline with the final
version of CalFUSE \citep{dix07} and further processed using our
suite of FORTRAN programs, unix scripts, and IRAF procedures
especially written for this purpose \citep{god12}.  
The {\it FUSE} data come in the form of eight spectral segments
(SiC1a, SiC1b, SiC2a, SiC2b, LiF1a, LiF1b, LiF2a, and LiF2b), which 
have to be combined together to give the final {\it FUSE} spectrum. 
These spectral segments do overlap and provide a way to renormalize 
the spectra in the SiC1, SiC2, and LiF2 channels to the flux in the 
LiF1 channel (which is the most reliable segment). In the present case, 
the two SiC channels experienced intermittent (and unexplained) drops 
in the count rates (which was more severe for the SiC2 channel than 
for the SiC1 channel). When combining the 8 spectral segments together, 
we took care to renormalize the SiC spectral segments to the LiF spectral segments.

The {\it FUSE} spectrum with line identifications is displayed in Fig.1. 
Many ISM molecular hydrogen absorption lines are indicated with vertical 
tick marks. Numerous helium lines dominate the shortest wavelengths. 
Also seen are neon emission lines. FPN is a fixed pattern noise from 
the {\it FUSE} detector. 

The {\it FUSE} spectrum was previously analyzed by \citet{eri06} 
who studied the locations, origins and excitation mechanisms of 
the emission lines, and by \citet{sko17} who used the {\it FUSE} spectrum 
to analyse the O\,{\sc vi} 1032~\AA\ line as part of a 
multi-wavelength analysis. 

In preparation for the spectral analysis, we dereddened the  {\it FUSE} 
spectrum assuming $E(B-V) = 0.10$ (see Table 1) and using our dereddening 
script based on the extinction curve from \citet{fit07} with $R=3.1$.

\begin{figure*}
\centering  
\includegraphics[scale=0.6]{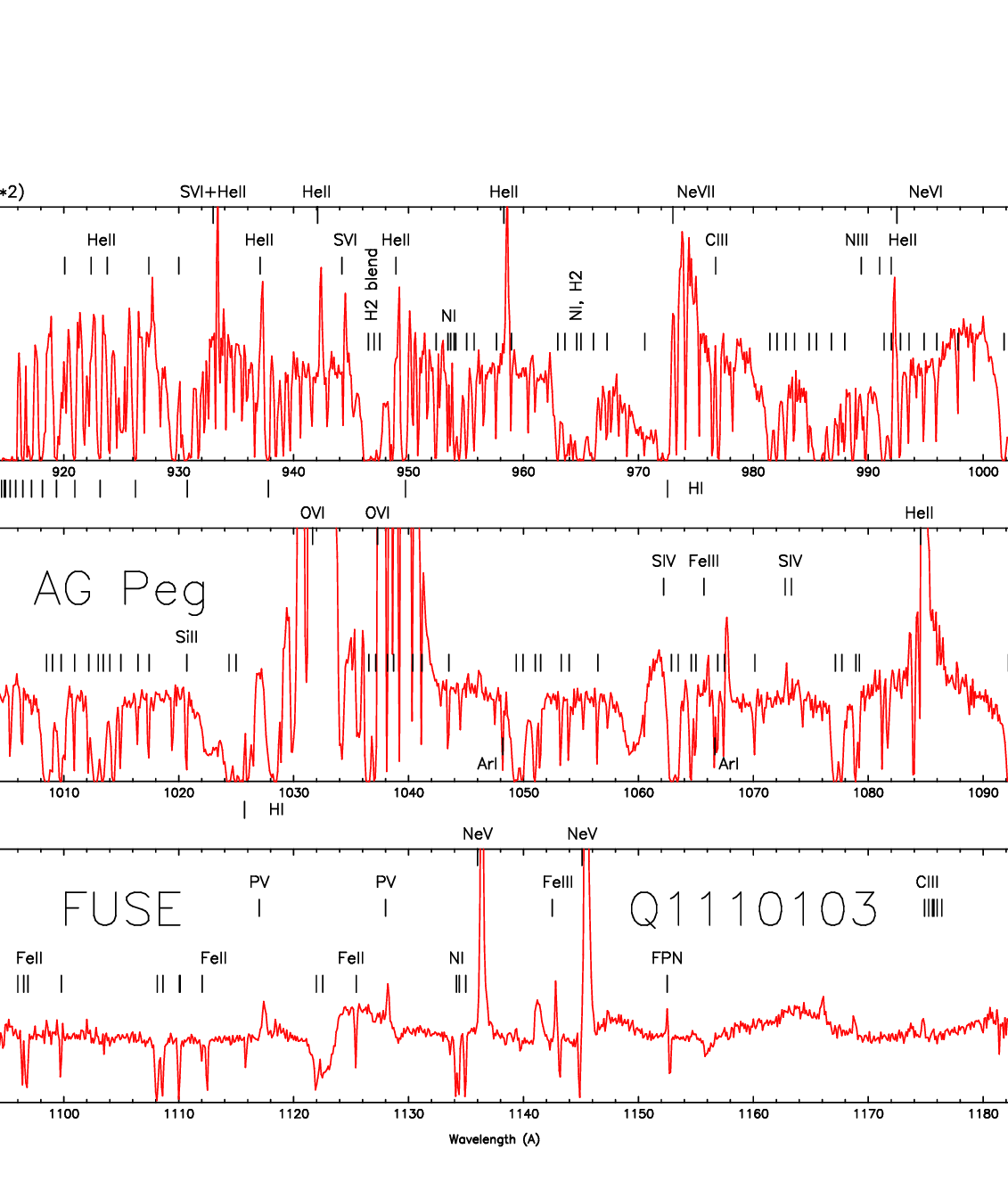}                 
\caption{
The {\it FUSE} spectrum of AG Peg is presented with line identifications.
The spectrum has not been dereddened.  
The spectrum is strongly affected by ISM absorption lines. The ISM molecular
hydrogen absorption lines are indicated with tick marks in the middle of 
each panel (just above the spectrum), the atomic hydrogen absorption
lines are indicated below each panel. Some additional sharp ISM absorption
lines are of iron, nitrogen, argon, and silicon. We identify sharp emission
lines of helium, neon, as well as sulfur and phosphorus. All the tick marks
have been placed at the rest wavelength and show that the sharp emission
lines are slightly red-shifted. The oxygen doublet (O\,{\sc vi}) lines
are the only lines in broad emission. The C\,{\sc iii} (1175) 
line has been marked but is not clearly detected. 
The "FPN" mark is a known detector fixed pattern noise.   
The lines in the {\it FUSE} spectrum of AG Peg have been studied in 
details in the work of \citet{eri06}.  
}
\end{figure*}

\clearpage

\subsection{Spectral Modeling and Analysis.} 

The spectral analysis is carried out by fitting the observed {\it FUSE} 
spectrum of AG Peg with theoretical (synthetic) model spectra 
of WDs and accretion disks. 

We use Hubeny's suite of codes TLUSTY and SYNSPEC \citep{hub88,hub95} 
to generate synthetic spectra for high-gravity stellar stmosphere 
WD models. A one-dimensional vertical stellar atmosphere structure is 
first generated with TLUSTY for a given surface gravity ($log(g)$),
effective surface temperature ($T_{\rm eff}$), and surface composition
(here we assume solar composition). The code SYNSPEC is then used to   
solve for the radiation field and generate a synthetic stellar spectrum
over a given wavelength range between 900~\AA\ and 7500~\AA . 
Finally the code ROTIN is used to reproduce rotational and instrumental
broadening as well as limb darkening. In this manner we generate a grid
of stellar photospheric models covering a wide range of effective temperatures
and surface gravities.    

The accretion disk spectra are generated by dividing the disk into rings,
each  with a given radius, temperature, effective surface (vertical) 
gravity, and density obtained from the standard disk model \citep{pri81}. 
For each ring the code TLUSTY is then run to generate a one-dimensional
vertical structure. This is then followed by a run of SYNSPEC 
to create a spectrum for each individual ring. The ring spectra are then
combined using DISKSYN which includes the effects of Keplerian rotational
broadening, inclination and limb darkening.     
All our disk models used here, as well as \citet{wad98}'s disk models,
have solar abundances.  

A complete description of our recently-updated accretion disk modeling is given in 
\citet{god17,dar17} and \citet{god18}. The main implementation 
in our disk models is that the inner and outer radii of the disk
are chosen to fit the parameters of the system that is being modeled. In the present case, the outer radius of the disk was extended
to several hundred times the radius of the accreting white dwarf, 
where the disk temperature drops to $\sim$ 3,500~K. In the disks 
of \citet{wad98}, the outer disk extends only to where the temperature
reaches 10,000~K. Nevertheless, because of the large mass accretion
rate (and ensuing higher disk temperature) and the short
wavelength coverage of {\it FUSE}, the inclusion of a more extended
(and colder) outer disk does not contribute additional flux to the FUV  
(our new disk spectra models also extend into the optical where the larger
outer disk significantly increases the flux). 

In the present work we also use model accretion disks from the optically thick, steady state disk model grid of \citet{wad98} as explained in the next section. 

Since the {\it FUSE} spectrum of AG Peg is heavily affected by ISM
molecular hydrogen absoprtion lines, we decided to model the ISM lines
to obtain a better fit. Since our main purpose is not to assess the
hydrogen atomic and molecular column densities, but only to improve 
our spectral fit, we use some of the ISM models we generated in 
\citet{god09} and refer the reader to that work for further details.

\clearpage 

\section{Results} 

In the following, we adopt a distance of 800 pc, an inclination of 
$50^{\circ}$, a WD mass of about $0.65 M_{\odot}$ and we expect
the WD temperature to be at least of the order of 100,000~K  
(see Table 1). We carry out all the disk and 
photosphere model fits for these values.

\subsection{NLTE WD Photosphere Analysis} 

The input gravity of the models was varied from $log(g)=6.0$ 
to $log(g)=8.0$ in steps of 0.5, and the WD temperature was varied
from $\sim 40,000$~K to $\sim 200,000$K (in steps of 5,000~K and 
10,000~K).  
We find that the best fit WD photosphere models are obtained for the 
lowest gravity ($log(g) = 6.0, 6.5$) in our grid of models and for a 
temperature $T_{\rm wd} =100,000$~K to 180,000~K. 
The radius of the WD is then obtained by scaling the theoretical 
spectrum obtained for a distance of 800~pc to the observed spectrum. 
The exact mass of the WD is uknown but it is likely 
around $\sim 0.65 M_{\odot}$ (see Table 1), and an output value of the gravity,
$log(g)$, can then be obtained and compared to the input value of $log(g)$
for self-consistency.  

Explicitly, for $T_{\rm wd} = 100,000$~K, the WD radius scaled to a distance 
of 800 pc is $R_{\rm wd}= 55.5 \times 10^9$cm ($\sim 0.08 R_{\odot}$), 
giving $log(g) = 6.44$ for an assumed WD mass of $M_{\rm wd}= 0.65M_{\odot}$ 
(see Table 2). Values of $log(g)$ are also listed for a $0.4M_{\odot}$ 
and $1 M_{\odot}$ WD masses and are consistant with the best fits obtained
for $log(g)=6.0$ and $log(g)=6.5$.  
 
For the hottest model $T_{\rm wd}=180,000$~K, the WD radius scaled 
to a distance of 800~pc is $R_{\rm wd}= 39.6 \times 10^9$cm 
($0.057 R_{\odot}$), giving $log(g) = 6.74$ for a $0.65 M_{\odot}$ WD
mass (Table 2). 
The 100,000~K model is slightly defficient in flux around 930-950~\AA , 
while the 180,000~K model provides slightly too much flux in that
region. At longer wavelengths, the models provide the same fit.  
Among these models, the best-fitting photosphere 
(namely in the 930-950~\AA\ region) has 
$T_{\rm wd}= 150,000$~K, and the WD radius scaled to a distance of 800~pc 
is $R_{\rm wd}= 42.9 \times 10^9$cm ($0.0615 R_{\odot}$), giving 
$log(g) = 6.67$ for a $0.65 M_{\odot}$ WD mass.
This WD photosphere model fit to the {\it FUSE} spectrum is displayed in Fig.2, 
and includes the addition of ISM absorption lines. We note that our 
derived value of $T_{\rm wd}$ agrees with the multi-wavelength analysis of 
\citet{sko17} who took $T_{\rm wd}= 160,000$~K but did not model the 
{\it FUSE} spectrum explicitly. The low mass (and/or low gravity) 
of the white dwarf is consistent 
with the very small mass function derived from the full orbit solution of 
\citep{fek00}. 

To further check the self-consistency of these models, we compute the  
luminosity (see Table 2, last column). Our best-fit model has a luminosity
of 1729$L_{\odot}$, which agrees with \citet{sko17}'s estimate, which
is not surprizing since the temperature is of the same order as 
\citet{sko17}'s. However, our hotest model has a luminosity about twice
as large which must be ruled out.   

\begin{deluxetable*}{ccccccc} 
\tablewidth{0pt}
\tablecaption{
Single hot component fitting results     
}
\tablehead{ 
 $T_h$   &  $R_h$    & $R_h$       & $M_{\rm h}$ & $Log(g)$  &  $L$  & $L$    \\ 
(1000 K) & (1000 km) & ($R_{\odot}$)  &($M_{\odot}$)&  (cgs) & ($L_{\odot}$) & (erg/s)    
}
\startdata
  100    &   55.5    &  0.0798     & 0.4         & 6.23  &   572   & $2.19 \times 10^{36}$  \\ 
  100    &   55.5    &  0.0798     & 0.65        & 6.44  &   572   & $2.19 \times 10^{36}$  \\ 
  100    &   55.5    &  0.0798     & 1.0         & 6.63  &   572   & $2.19 \times 10^{36}$  \\ 
  150    &   42.9    &  0.0615     & 0.4         & 6.46  &   1729  & $6.64 \times 10^{36}$  \\ 
  150    &   42.9    &  0.0615     & 0.65        & 6.67  &   1729  & $6.64 \times 10^{36}$  \\ 
  150    &   42.9    &  0.0615     & 1.0         & 6.86  &   1729  & $6.64 \times 10^{36}$  \\ 
  180    &   39.6    &  0.0570     & 0.4         & 6.53  &   3056  & $1.17 \times 10^{37}$  \\ 
  180    &   39.6    &  0.0570     & 0.65        & 6.74  &   3056  & $1.17 \times 10^{37}$  \\ 
  180    &   39.6    &  0.0570     & 1.0         & 6.93  &   3056  & $1.17 \times 10^{37}$  \\ 
\enddata
\tablecomments{
The radius of the hot component $R_h$ is derived by scaling
the model flux to the observed flux assuming a distance of 800~pc.  
The output gravity is then obtained assuming a given mass 
for the hot component. 
}    
\end{deluxetable*}

\clearpage

\begin{figure*}
\centering  
\includegraphics[scale=0.6]{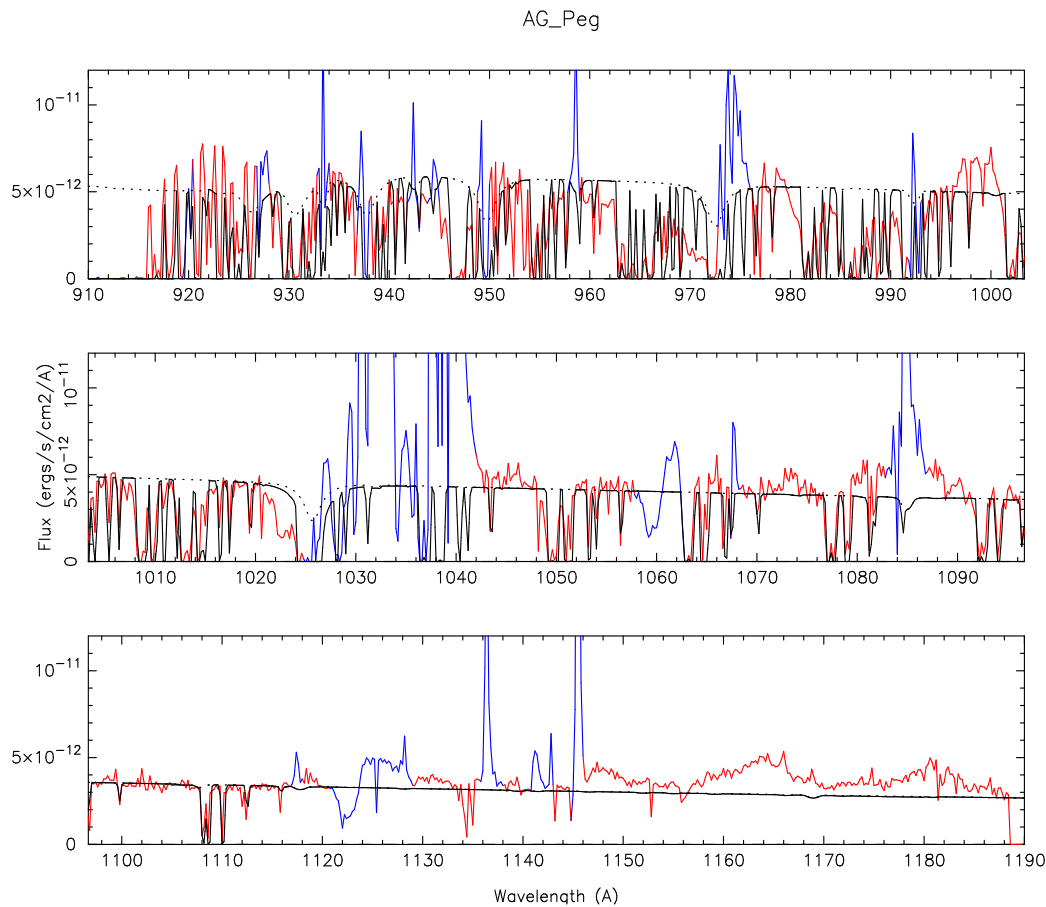}                 
\caption{
Single WD fit to the spectrum of AG Peg. 
The {\it FUSE} spectrum has been dereddened assuming
$E(B - V ) = 0.10$ and it is drawn in red, regions that have strong emission 
lines and possible P-Cygni profiles have
been masked and are in blue. The synthetic stellar spectrum model 
(solid black line) has a temperature of 150,000~K,
gravity $log(g) = 6.00$, solar abundances, projected rotational velocity 
of 200 km/s, and include an ISM absorption model. 
The stellar spectrum model without the ISM absorption is shown with the 
dashed black line.}
\end{figure*}

\clearpage 

\subsection{NLTE Accretion Disk Analysis}

We extended our analysis of the {\it FUSE} spectrum to optically thick, 
steady state accretion disk models. We first used the grid of UV 
disk models by \citet{wad98}, and found out that AG Peg must have 
a high mass transfer rate if one ascribes all of the FUV flux to 
accretion luminosity. We tried the highest mass accretion rate models of
\citet{wad98} with $\dot{M}= 10^{-8}M_{\odot}$/yr, $M_{\rm wd}= 0.55M_{\odot}$ 
and $M_{\rm wd}= 0.80M_{\odot}$, all with an inclination $i = 41^{\circ}$.
However, all these disk models yielded derived distances much shorter, 
and did not provide enough flux in the short wavelength range of {\it FUSE}  
($\lambda  < 960$~\AA). When linearly scaling these disk models 
to 800 pc , we found that the mass accretion rate should 
be of the order of $\dot{M} \approx 10^{-7}-10^{-6}M_{\odot}$/yr. 
Consequently, we generated accretion disk models for 
larger mass accretion rates using TLUSTY and SYNSPEC \citep{hub88,hub95} 
as described in the previous section.

We chose a WD mass $M_{\rm wd}= 0.7M_{\odot}$, with a  
radius $R_{\rm wd}=8,500$~km, and an inclination $i = 50^{\circ}$. 
The disk extends to where its temperature $T_{\rm disk}$ drops to 
$\approx 3,500$~K.
For a given mass accretion rate the disk model scales to a given distance.

To match a distance of 800 pc we obtained that the mass accretion rate 
has to be $\dot{M} = 1.8 \times 10^{-7}M_{\odot}$/yr. 
This disk model has an inner radius of 8,500 km 
and an outer radius of 1 million km, and is presented in Figure 3. 
The fit in the shorter wavelength of {\it FUSE} is not as good as for the
single hot WD model fit (Figure 2), as the model is deficient in flux.   

We note that even if we increase the mass accretion to 
$\dot{M} = 1 \times 10^{-6}M_{\odot}$/yr, the fit in the short
wavelength range is still not as good as the single WD component. 
This model (shown  in Figure 4) leads a distance of 1320 pc. 
 
We also tried combined WD + accretion 
disk models where the WD contributes a significant fraction of the
flux, and the result was intermediate between the single WD models 
and the single disk models. Namely, the WD+disk models were improved 
over the single disk models, but did not provide a fit as good as the 
single WD models.

\clearpage

\begin{figure*}
\centering  
\includegraphics[scale=0.6]{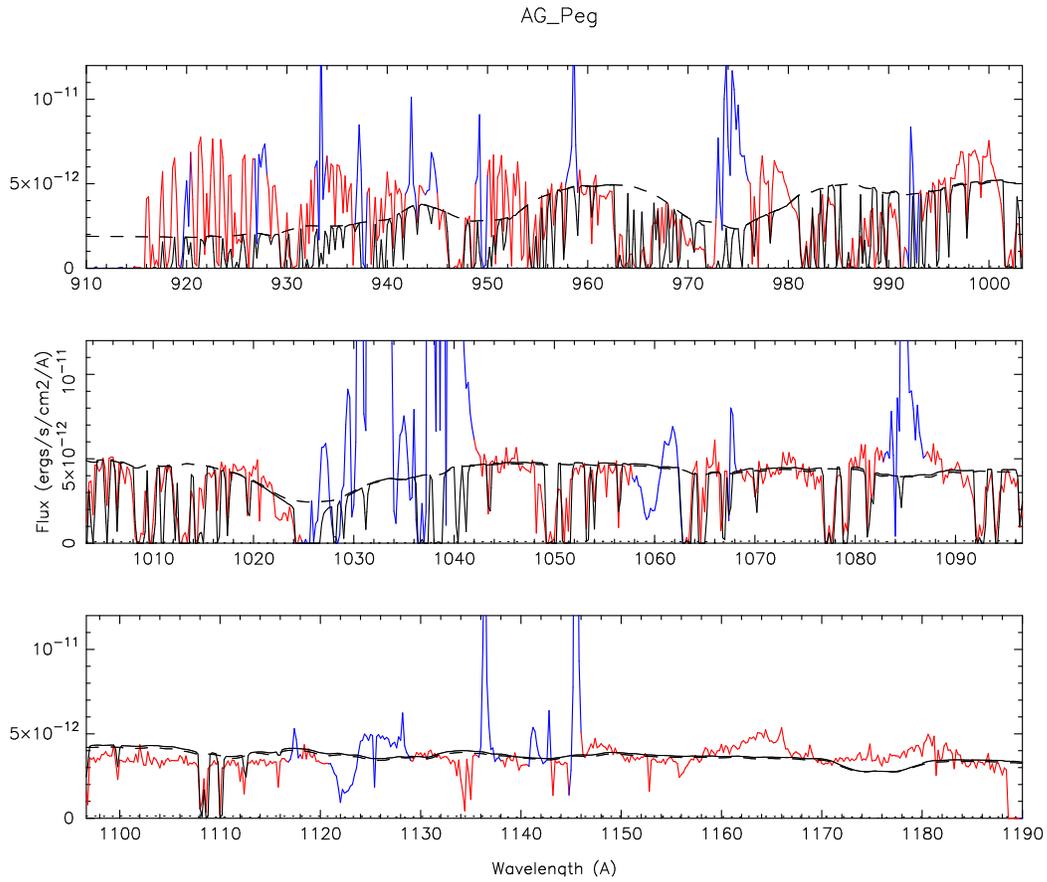}                 
\caption{Accretion Disk model fit to the {\it FUSE} spectrum of AG Peg. The accreting WD mass has been set to
$M_{\rm wd}= 0.7M_{\odot}$, with a radius $R_{\rm wd}=8,500$~km. 
The disk model has a mass accretion rate 
$\dot{M} = 1.8 \times 10^{-7}M_{\odot}$/yr, 
an inner radius of 8,500 km, an outer radius of
$\sim$1 million km, with  solar abundances. The
inclination of the system has been set to $i = 50^{\circ}$. 
The {\it FUSE} spectrum is in red (with masked portions in blue), and
the disk model (including ISM absorption) is drawn with the solid black line (the model without the ISM absorption
is drawn with the dashed black line).}
\end{figure*}

\clearpage

\begin{figure*}
\centering  
\includegraphics[scale=0.6]{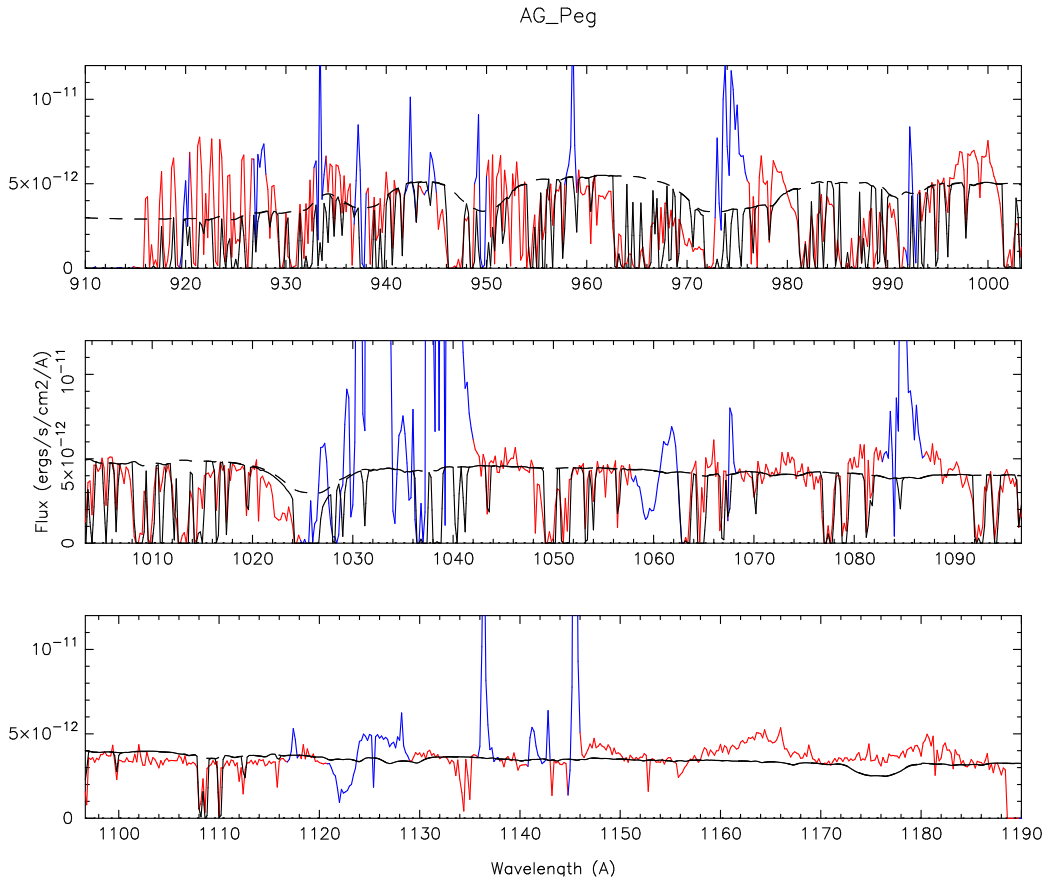}                 
\caption{Accretion Disk model fit to the {\it FUSE} spectrum of AG Peg. 
The accreting WD mass has been set to
$M_{\rm wd}= 0.7M_{\odot}$, with a radius $R_{\rm wd}=8,500$~km. The disk model has an inner radius of 9,350 km, an outer radius of
4 million km, a mass accretion rate $\dot{M} = 1.0 \times 10^{-6}M_{\odot}$/yr, and solar abundances. The
inclination of the system has been set to $i = 50^{\circ}$. 
The {\it FUSE} spectrum is in red (with masked portions in blue), and
the disk model (including ISM absorption) is drawn with the solid black line (the model without the ISM absorption
is drawn with the dashed black line).}
\end{figure*}

\section{Discussion and Conclusions} 

On the basis of fitting very hot NLTE white dwarf photosphere models 
to the {\it FUSE} spectrum of AG Peg in quiescence and comparing
with the best fitting high 
$\dot{M}$ ($10^{-7}-10^{-6}M_{\odot}$/yr) disk models, 
we report evidence that a very hot white dwarf with a temperature 
$T_{\rm wd}$ = 150,000K dominates the entire {\it FUSE} wavelength 
range while an accretion disk with high $\dot{M}$ gives a 
poorer fit in the short wavelengths ($  \lambda <$ 960~\AA).  
This hot WD model 
has a gravity $\log(g)\sim 6.5$, and the distance of 800 pc gives
a scaled radius $R_{\rm wd}\sim 43 \times 10^{9}$ cm ($\sim 0.06 R_{\odot}$).  
The addition of an accretion disk only degrades the hot WD fit and implies 
that the WD is possibly over-shining the disk.

It is interesting to compare our results on the hot component of AG Peg using 
{\it FUSE} spectra with a recent multi-wavelength analysis  
by \citet{sko17}. 
They computed the peak luminosity of the June 2015 Z And-like
outburst of AG Peg to be $2 - 11 \times 10^{37}$ergs/s (for a distance of 0.8 kpc). 
They estimated that  the white dwarf had to be accreting at 
$\sim 3 \times 10^{-7}M_{\odot}$/yr, which they claimed exceeds 
the stable burning limit by which one assumes is the steady state 
limit in which the accreted material is burned at the same rate that 
it accretes. 

Our best-fitting NLTE high gravity photosphere models were fit best with 
$T_{\rm wd} = 150,000$~K which is in agreement with the modeling of the
UV/Optical continuum in quiescence from 1993 and 1996 \citet{sko17}, and  
considerably lower than the hot component of AG Peg in outburst.
The luminosity of our best-fit model also agrees with the quiescent luminosity
estimate of \citet{sko17}. 
During the outburst, the effective radius of the hot component, $R_h$, varies
between $\sim 0.09R_{\odot}$ and $\sim 0.15R_{\odot}$. 
The accretion rate they found, on average, was a factor of 2-3 larger during the outburst than during quiescence. 

Note that \citet{mur91} obtained temperatures and luminosities of the 
hot component in symbiotic novae using the modified Zanstra Method. 
For AG Peg they obtained a hot component temperature between 95,000K 
and 100,000K during the {\it decline phase}.
However, we note that
\citet{mur91} adopted a shorter distance (650 pc vs 800 pc) as well as a 
smaller E(B-V) = 0.05 (vs 0.10) which leads to a lower
effective temperature for the hot component (since the UV continuum is less steep
and a much lower luminosity than that derived by \citet{ken93,ken01}). 
Other attempts assumed black bodies to fit the SED of 
AG Peg encompassing the optical all the way to the IUE SWP range. 
These efforts obtained temperatures significantly cooler than the 
temperature of 150,000~K that we obtained from
the best fitting NLTE model atmosphere to the {\it FUSE} spectrum. 

It is interesting to note that our method of deriving the temperature
of the hot component gives a temperature ($Log(T_{\rm eff})=5.17$) 
similar to the temperature obtained from an analysis of the optical 
H\,{\sc i} and He\,{\sc ii} lines \citep[see Fig.5 in ][]{ken01},   
but much higher than the temperature derived from
an analysis of the UV lines or the UV continuum.   
The luminosity we derive here is significantly larger than
derived by these 3 methods \citep{ken01}.  
In the present work we dereddened the {\it FUSE} spectrum using 
the extinction law given by \citet{fit07}, while
\citet{ken01} adopted  \citet{mat90}'s extinction curve (which itself
is the \citet{car89} extinction law). 
In the FUV region the difference  between \citet{car89} and \citet{fit07}
extinction laws increases with decreasing wavelength and can gives
different results \citep[see e.g. ][]{sel13}. 
It is possible that the difference in the extinction laws in the shortest
wavelengths of {\it FUSE} increased the flux (in the dereddened spectrum)
to better match a 150,000~K model, since the difference between the
100,000~K model and the 150,000~K model is very small. 
A temperature closer to 100,000~K gives a luminosity in better 
agreement with the quiescent luminosity.       

It is also possible that the observed difference in temperature is due to a real
change in the WD surface temperature. 
The short wavelength range of the {\it FUSE} spacecraft down to the Lyman 
Limit probes  the radiation from the innermost disk/boundary 
layer/white dwarf, which likely accounts for the 
higher temperature that we obtain. No previous attempts to derive 
a temperature for the hot component in AG Peg covered a wavelength 
range down to the Lyman Limit.

Overall, our result suggests the possibility that 
the system is dominated by 
a very hot, low gravity white dwarf. 
This is  
similar to our analysis of the {\it FUSE} spectrum of the S-Type 
symbiotic star RW Hydrae \citep{sio17}, which indicates it
also contains a very hot, bare white dwarf 
with no evidence of an accretion disk. An important question now 
is whether the Z And-type outbursts are due to bursts of accretion from the 
M3/4 III companion or if they 
are also thermonclear powered like the 1850 nova outburst.

On the other hand, the actual structure of an accretion disk 
surrounding a hot stellar component in symbiotic binary
is poorly understood compared to catalcysmic variables. 
First, the scale of a symbiotic binary relative 
to a cataclysmic variable is vastly larger and second, the mass transfer 
occurs via a red giant donor which may or may not fill its
Roche lobe. This points to the possibility that the disk models 
might be different than the standard disk model. 
3D hydrodynamic model simulations of symbiotic binaries 
\citep{dev09,dev17} reveal 
that a disk-like structure resembling a steady-state disk does appear to form, 
but the thin disk approximation may break down in symbiotics. 
Given the uncertainty in what a true symbiotic accretion 
disk would look like, and that a very hot low gravity degenerate star dominates the FUV flux,
the presence of a steady-state (standard)
accretion disk cannot be summarily ruled out nor denied.

However, we note that for the hot WD to contribute a significant fraction of 
the total flux to the {\it FUSE} spectrum, in addition to that
of an accretion disk, its emitting photosphere 
radius has to be inflated (Section 3.1). 
With a radius of the order of $R_{\rm wd} \sim 0.05 R_{\odot}$,  
rather than $R_{\rm wd} \sim 0.01 R_{\odot}$,   
the accretion disk temperature drops significantly \citep{god17}. For example, 
at a mass accretion rate of $\dot{M}=10^{-7}M_{\odot}$/yr
the peak temperature in the disk reaches 103,000~K if $R_{\rm wd}=8,500$~km
(which is the radius of a $\sim 30,000$~K WD of mass $0.7M_{\odot}$).  
However, for a WD radius of $R_{\rm wd}=35,000$~km $ \approx 0.05 R_{\odot}$,
the peak temperature of such a disk drops to 36,500~K.  
At a mass accretion rate of $\dot{M}=10^{-6}M_{\odot}$/yr, the disk peak
temperature drops from $\sim$171,000~K to $\sim 65,000$~K.  
In both cases ($\dot{M}=10^{-7}M_{\odot}$/yr and $\dot{M}=10^{-6}M_{\odot}$/yr)
the disk will contributes little flux to the {\it FUSE}  
spectrum if $R_{\rm wd}=0.05 R_{\odot}$.
As the accretion luminosity decreases like $\propto 1/ R_{\rm wd}$,
so does the disk luminosity, and as long as the WD radius remains inflated
it will dominate the FUV spectrum.

\acknowledgements 
This work is supported by funding from the National Aeronautics and Space Administration (NASA) under grant number NNX17AF36G issued through the Office of Astrophysics and DATA Analysis Program (ADAP) to Villanova University. This study has also been supported in part by the National Science Centre, Poland, grant OPUS 2017/27/B/ST9/01940. 
PG is pleased to thank William (Bill) P. Blair at the 
Henry Augustus Rowland Department of Physics \& Astronomy at the 
Johns Hopkins University, Baltimore, Maryland, USA, for his kind hospitality. We made use of online data from the AAVSO and we are thankful to the AAVSO and its members worldwide for their constant monitoring of CVs and for making their data public.   

\facilities{{\it FUSE}}

\software{
IRAF (v2.16.1, \citet{tod93}), 
Tlusty (v203) Synspec (v48) Rotin(v4) Disksyn (v7) \citep{hub17a,hub17b,hub17c}, 
PGPLOT (v5.2), Cygwin-X (Cygwin v1.7.16),
xmgrace (Grace v2), XV (v3.10) } 
\\ \\

Patrick Godon \url{https://orcid.org/0000-0002-4806-5319} 
 
Joanna Mikolajewska \url{https://orcid.org/0000-0003-3457-0020}

\end{document}